\documentclass[showpacs,twocolumn]{revtex4}
\usepackage{graphicx}

\begin{document}

\title{Comment on ``On the tunneling through the black hole horizon''}

\author{S. A. Hayward$^a$,
R.~Di Criscienzo$^{a,b}$, M.~Nadalini$^b$, L.~Vanzo$^b$, S.~Zerbini$^b$}

\address{$^a$Center for Astrophysics, Shanghai Normal University, 100
Guilin Road, Shanghai 200234, China}

\address{$^b$Dipartimento di Fisica, Universit\`a di Trento and INFN,
Gruppo Collegato di Trento, Italia}

\date{22nd November 2009}

\begin{abstract}
The arguments of the above article do not apply to the papers which it 
criticizes, and contain several key errors, including a fundamental 
misunderstanding about the equivalence principle. 
\end{abstract}
\pacs{04.70.-s, 04.70.Bw, 04.70.Dy} 
\maketitle 

\begin{figure}
\includegraphics[width=3cm]{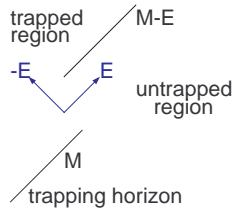}
\caption{
Space-time diagram of pair production just inside a Schwarzschild black hole of 
mass $M$, with an outgoing particle of positive energy $E$ and an ingoing 
particle of negative energy $-E$: including back-reaction, the black-hole mass 
is reduced to $M-E$ and so its trapping horizon shrinks, allowing the outgoing 
particle to escape.} \label{tun} 
\end{figure}

Hawking radiation from black holes \cite{Haw} is a well established prediction 
of quantum field theory on curved space-times 
\cite{dewitt,BD,wald,fulling,visser,igor}. The intuitive picture of tunneling 
out of the black hole was made precise by Parikh \& Wilczek \cite{PW}, and can 
be summarized in a simple diagram (Fig.\ref{tun}). This method used a 
quasi-stationary or adiabatic approximation. In full quantum field theory, the 
trapping horizon shrinks gradually, becoming time-like so that escape from the 
trapped region is possible. While many derivations of Hawking temperature do 
not obviously generalize beyond stationary cases, we have recently shown that a 
Hamilton-Jacobi variant of the tunneling method can be so generalized, yielding 
a local temperature for a trapping horizon \cite{CNVZZ,HCNVZ,CHNVZ}. 

Sadly there are a tiny but vociferous minority who do not believe in such 
tunneling at all, and instead believe that they have disproved it already. The 
latest such foray by Belinski \cite{Bel}, which cites \cite{PW,CNVZZ,CHNVZ} 
among others, is the subject of this Comment. We begin by noting that the paper 
does not allege any specific error in the quite straightforward calculations, 
e.g.\ \cite{HCNVZ}, but instead makes arguments of principle as to why the 
effect is impossible, which turn out to be groundless if one actually makes 
relevant calculations, as follows. 
\begin{enumerate}

\item The author begins with the statement that the ``argumentation of all 
    these papers are as following'', then goes through a calculation for the 
    Schwarzschild black hole in the usual $(t,r)$ coordinates. We did not use 
    those coordinates in \cite{CNVZZ,HCNVZ,CHNVZ}, since they break down at the 
    horizons and so are inadmissable. 

\item The author argues that classical trajectories cannot exit through the 
    future horizon. But we are doing quantum physics, allowing for 
    back-reaction, as above. 

\item The author argues further about trajectories, which are irrelevant; 
    the method \cite{CNVZZ,HCNVZ,CHNVZ} derives a temperature which is a 
    property of a section of the trapping horizon, independent of trajectory. 

\item The author claims that the calculations might apply to the past 
    horizon rather than the future horizon, which is the opposite of the 
    truth. For instance, the derivation in \cite{HCNVZ} used advanced 
    Eddington-Finkelstein coordinates, which break down at past horizons 
    but cover future horizons, where the temperature was derived. 

\item The author claims that the effect is impossible by {\em ``the 
    equivalence principle''}, since in a {\em ``locally inertial system the 
    equations of geodesics and Hamilton-Jacobi equation for the action are the 
    same as in flat Minkowski space-time''} (his italics). This reveals a 
    fundamental misunderstanding of the equivalence principle. The most that 
    such an argument could show is that, at a given space-time point, there 
    exist observers, necessarily inertial, for whom the method yields no 
    temperature. In the simplest example of a Schwarzschild black hole, the 
    static observers are not inertial but accelerating, in order to remain 
    static in the gravitational field, as required by {\em the equivalence 
    principle} (our italics). They do measure a non-zero temperature, as is 
    well known \cite{Haw,dewitt,BD,wald,fulling,visser,igor} and confirmed by 
    the Hamilton-Jacobi method \cite{CNVZZ,HCNVZ,CHNVZ}. 

\item The final paragraph claims some problem with space-times which are 
    non-static, or apparently non-static in given coordinates. Our version of 
    the Hamilton-Jacobi method \cite{CNVZZ,HCNVZ,CHNVZ} applies to non-static 
    space-times. 
\end{enumerate}

\medskip
SAH was supported by the National Natural Science Foundation of China under
grants 10375081, 10473007 and 10771140, by Shanghai Municipal Education
Commission under grant 06DZ111, and by Shanghai Normal University under grant
PL609.


\begin{thebibliography}{99}
\bibitem{Haw}S W Hawking, Nature {\bf 248}, 30 (1974); {Commun. Math. Phys.}
    {\bf 43}, 199 (1975) [Erratum-ibid.\  {\bf 46}, 206 (1976)]
\bibitem{dewitt}
  B.~S.~DeWitt,
  Phys.\ Rept.\  {\bf 19} (1975) 295.
\bibitem{BD}N D Birrell \& P C W Davies, Quantum fields in curved space
    (Cambridge University Press 1982).
    {\bf 43}, 199 (1975) [Erratum-ibid.\  {\bf 46}, 206 (1976)]
\bibitem{wald}R M  Wald, Quantum Field Theory in Curved Spacetime and
    Black Hole Thermodynamics 
    (Chicago Lectures in Physics,  Chicago University Press   1994).
\bibitem{fulling} S. A. Fulling, Aspects of Quantum Field Theory in
    Curved Space-time (Cambridge  University Press   1996).
\bibitem{visser}
  M.~Visser,
  Int.\ J.\ Mod.\ Phys.\  D {\bf 12}, 649 (2003)
\bibitem{igor} V.~P.~Frolov and I.~D.~Novikov, Black hole physics, Kluwer 
    Academic Publisher, 2007. 
\bibitem{PW} M.~K.~Parikh and F.~Wilczek, Phys.\ Rev.\ Lett.\  {\bf 85}, 5042  
    (2000). 
\bibitem{CNVZZ}
  R.~Di Criscienzo, M Nadalini, L Vanzo, S Zerbini and G Zoccatelli,
  Phys.\ Lett.\  B {\bf 657}, 107 (2007).
\bibitem{HCNVZ}S A Hayward, R Di Criscienzo, M Nadalini, L Vanzo \& S Zerbini,
    {Class. Quantum Grav.} {\bf 26}, 062001 (2009)
\bibitem{CHNVZ}R Di Criscienzo, S A Hayward, M Nadalini, L Vanzo \& S Zerbini,
 Invariance of the Tunnelling Method for Dynamical Black Holes,
 arXiv:0906.1725 
\bibitem{Bel}V Belinski, On the tunneling through the black hole horizon, 
    arXiv:0910.3934 
\end{thebibliography}
\end{document}